\documentclass{PoS}

\usepackage{amsmath}
\usepackage{amssymb}
\usepackage{bm} 
\usepackage{cite}

\newcommand{\del}{\partial}
\newcommand{\nn}{\nonumber}

\title{Status of Chiral-Scale Perturbation Theory}

\ShortTitle{Chiral-Scale Perturbation Theory}

\author{R.~J.~Crewther \\
 CSSM and ARC Centre of Excellence for Particle Physics at the Tera-scale, \\
	  Department of Physics, University of Adelaide, \\
	  Adelaide, South Australia 5005, Australia \\
       E-mail: \email{rcrewthe@physics.adelaide.edu.au}}

\author{\speaker{Lewis~C.~Tunstall} \\
       Albert Einstein Center for Fundamental Physics, \\
       Institute for Theoretical Physics,  University of Bern, \\
       Sidlerstrasse 5, CH--3012 Bern, Switzerland \\
       E-mail: \email{tunstall@itp.unibe.ch}}
       
\abstract{Chiral-scale perturbation theory $\chi$PT$_\sigma$ has been proposed 
as an alternative to chiral $SU(3)_L\times SU(3)_R$ perturbation theory which 
explains the $\Delta I = 1/2$ rule for kaon decays. It is based on a 
low-energy expansion about an infrared fixed point in three-flavor QCD.  In 
$\chi$PT$_\sigma$, quark condensation $\langle\bar q q \rangle_\mathrm{vac} 
\neq 0$ induces nine Nambu-Goldstone bosons: $\pi, K, \eta$ and a QCD dilaton 
$\sigma$ which we identify with the $f_0(500)$ resonance. Partial conservation 
of the dilatation and chiral currents constrains low-energy constants which 
enter the effective Lagrangian of $\chi$PT$_\sigma$.  These constraints allow 
us to obtain new phenomenological bounds on the dilaton decay constant 
via the coupling of $\sigma/f_0$ to pions, whose value is known precisely 
from dispersive analyses of $\pi\pi$ scattering.  Improved predictions for 
$\sigma \to \gamma \gamma$ and the $\sigma NN$ coupling are 
also noted. To test $\chi$PT$_\sigma$ for kaon decays, we revive a 1985 
proposal for lattice methods to be applied to $K \to \pi$ \emph{on-shell}.}

\FullConference{The 8th International Workshop on Chiral Dynamics, CD2015 \\
		29 June 2015 - 03 July 2015 \\
		Pisa, Italy}

\begin{document}

\section{Approximate Scale Invariance in Low-Energy QCD}
\begin{figure}[ht]           
\centering
\includegraphics[scale=0.8]{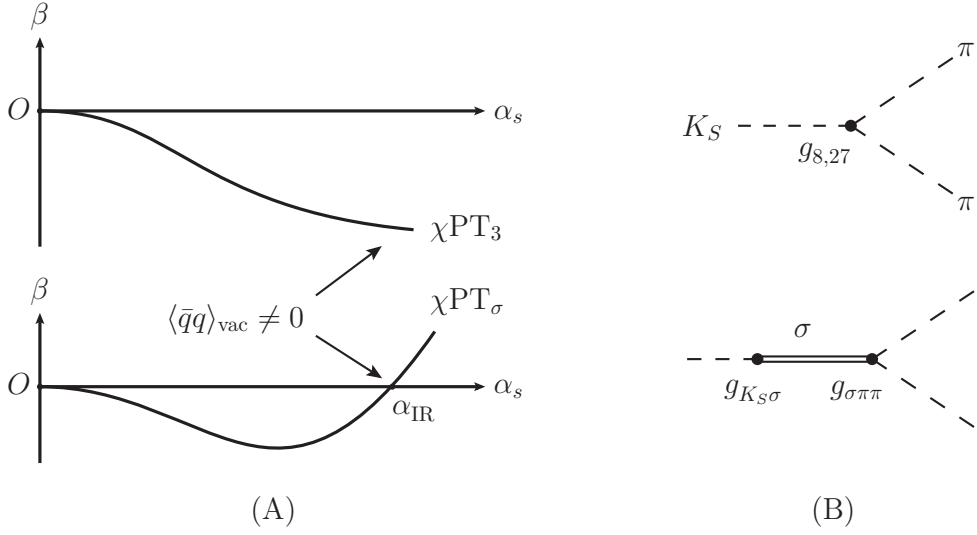}
\caption{(A) Scenarios for the $\beta$ function in three-flavor QCD, with 
corresponding low-energy expansions. In the absence of an infrared fixed point 
$\alpha^{}_\mathrm{IR}$ (top diagram), there is no approximate scale invariance 
and chiral $SU(3)_L\times SU(3)_R$ perturbation theory $\chi$PT$_3$ is relevant
at low-energies.  If $\alpha^{}_\mathrm{IR}$ exists (bottom diagram), quark 
condensation $\langle \bar q q \rangle_\mathrm{vac} \neq 0$ implies that the NG 
spectrum contains a QCD dilaton $\sigma$, and $\chi$PT$_3$ \emph{must} 
be replaced by chiral-scale perturbation theory $\chi$PT$_\sigma$. (B) Diagrams
for $K \to \pi\pi$ decay in lowest-order $\chi$PT$_\sigma$. The dilaton pole 
diagram is responsible for the dominant $\Delta I = 1/2$ amplitude.}
\label{fig:beta chpt}
\end{figure}

In the low-energy regime of QCD with heavy quarks $t,b,c$ decoupled, the 
relevance of scale (dilatation) invariance is determined by the trace 
anomaly~\cite{Adler77,Mink76,Niels77,Coll77} of the resulting 3-flavor theory:%
	\footnote{Here, $G_{\mu\nu}^a$ is the gluon field strength, 
  $\alpha_s = g_s^2/4\pi$ is the strong running coupling, and 
  $\beta = \mu^2 \del\alpha_s/\del\mu^2$ and 
  $\gamma_m = \mu^2\del \ln m_q/\del\mu^2$ refer to a mass-independent 
  renormalization scheme with scale $\mu$.}
\begin{equation}
  \theta^\mu_\mu =\frac{\beta(\alpha_{s})}{4\alpha_{s}} G^a_{\mu\nu}G^{a\mu\nu}
  + \bigl(1 + \gamma_{m}(\alpha_{s})\bigr)\sum_{q=u,d,s} m_{q}\bar{q}q\,.
  \label{anomaly}
\end{equation}
Depending on the infrared behaviour of $\beta$, there are only two 
realistic scenarios (Fig.\ \ref{fig:beta chpt} (A)):
\begin{enumerate}
\item If $\beta$ remains negative and non-zero, possibly diverging
linearly at large $\alpha_s$, scale invariance is explicitly broken by 
$\theta_\mu^\mu$ being large \emph{as an operator}. There is \emph{no hint} 
of approximate scale invariance: quantities such as the nucleon mass $M_N = 
\langle N | \theta_\mu^\mu | N \rangle$ are generated almost entirely by 
the gluonic term in (\ref{anomaly}). Then conventional chiral 
$SU(3)_L\times SU(3)_R$ perturbation theory $\chi$PT$_3$ is the appropriate 
low-energy effective theory for QCD amplitudes expanded in powers 
of $O(m_K)$ external momenta and light quark masses $m_{u,d,s}=O(m_K^2)$. 

\item If $\beta$ vanishes when $\alpha_s$ runs non-perturbatively 
to an infrared fixed point $\alpha^{}_\mathrm{IR}$, the gluonic term 
$\sim G_{\mu\nu}^a G^{a\mu\nu}$ in (\ref{anomaly}) is absent and the dilatation 
current $D_\mu = x^\nu \theta_{\mu\nu}$ becomes conserved in the limit of 
vanishing quark masses:
\begin{align}  
\left.\del^\mu D_\mu\right|_{\alpha_s = \alpha^{}_{\mathrm{IR}}} 
  =   \left.\theta^\mu_\mu\right|_{\alpha_s = \alpha^{}_{\mathrm{IR}}}
  &=  \bigl(1 + \gamma_{m}(\alpha^{}_{\mathrm{IR}})\bigr) 
      \sum_{q=u,d,s}m_q\bar{q}q \nonumber \\
  &\to 0\ , \ SU(3)_L\times SU(3)_R \mbox{ limit}\,.	
\label{scale}
\end{align}
Although the Hamiltonian preserves dilatations in this limit, 
\emph{the vacuum state is not scale invariant} due to the formation of a quark 
condensate $\langle \bar q q \rangle_\mathrm{vac} \neq 0$. As a result, both 
chiral $SU(3)_L\times SU(3)_R$ and scale symmetry are realized in the 
Nambu-Goldstone (NG) mode and the spectrum contains nine massless bosons: 
$\pi, K, \eta$ and a $0^{++}$ QCD dilaton $\sigma$.   Non-NG bosons remain massive 
\emph{despite the vanishing of} $\theta_\mu^\mu$ and have their scale set by  
$\langle \bar q q \rangle_\mathrm{vac}$.  The relevant low-energy expansion 
involves a combined limit
\begin{equation}
m_{u,d,s} \sim 0 \quad \mbox{and} \quad \alpha_s \lesssim \alpha^{}_\mathrm{IR}\,,
\end{equation}
and leads to a new effective theory $\chi$PT$_\sigma$ of approximate chiral-scale 
symmetry~\cite{CT1,CT2}.  In this scenario, the dilaton mass is set by $m_s$, so 
the natural candidate for $\sigma$ is the $f_0(500)$ resonance, a broad $0^{++}$ 
state whose complex pole mass has real part $\lesssim m_K$~\cite{Cap06,Kam11,PDG}.
\end{enumerate} 
Until now, scenario 1 has been the generally accepted view, but we 
have observed~\cite{CT1,CT2} that $\chi$PT$_\sigma$ offers several 
advantages over $\chi$PT$_3$: it explains the mass and width of $f_0(500)$, 
produces convergent chiral expansions as a result of $\sigma/f_0$ being
promoted to the NG sector, and most importantly, explains the 
$\Delta I =1/2$ rule for non-leptonic $K$ decays 
(Fig.~\ref{fig:beta chpt} (B)).

Because approximate scale symmetry is included, the effective Lagrangian for 
$\chi$PT$_\sigma$ (Sec.~\ref{Lag}) contains several new low-energy constants 
(LECs) yet to be determined precisely from data.  Of particular interest is the 
dilaton decay constant $F_\sigma$ given by 
$m_\sigma^2F_\sigma = -\langle\sigma|\theta^\mu_\mu|\text{vac}\rangle$. 
If $F_\sigma$ is roughly 100 MeV, scale breaking by the vacuum can generate 
large masses such as $m_N \approx F_\sigma g_{\sigma NN}$ (Goldberger-Treiman 
relation for dilatons~\cite{MGM62}) for $m_\sigma$ small. The imprecise value of 
$F_\sigma$ in our previous work~\cite{CT1,CT2} arose from large uncertainties 
in the phenomenological value of $g_{\sigma NN}$~\cite{CC08,CC10}.

We circumvent this difficulty in Secs.~\ref{tensor} and \ref{predict}. First,
we find new constraints on LECs in the $\chi$PT$_\sigma$ effective Lagrangian
by requiring full consistency with the dilatation and chiral currents being 
conserved in the limit (\ref{scale}). These constraints allow us to determine 
$F_\sigma$ from the $\sigma\pi\pi$ coupling, whose value is known to remarkable 
precision from dispersive analyses~\cite{Cap06,Kam11,PDG} of $\pi\pi$ scattering.  
Then we obtain improved predictions for the non-perturbative Drell-Yan ratio 
\begin{equation} 
R = \sigma(e^{+}e^{-}\rightarrow\mathrm{hadrons})/ 
\sigma(e^{+}e^{-}\rightarrow\mu^{+}\mu^{-})
\quad\mbox{at } \alpha^{}_\mathrm{IR}\,,
\label{RIR}
\end{equation} 
as well as the $\sigma NN$ coupling.

In Sec.~\ref{lattice}, we resurrect an old proposal~\cite{RJC86} 
to apply lattice QCD for $K\to\pi$ \emph{on-shell} to determine the
couplings $g^{}_{8,27}$ in Fig.~\ref{fig:beta chpt} (B). Comments 
on the validity of $\chi$PT$_\sigma$ are reviewed in Sec.~\ref{issues}.

\section{Chiral-Scale Lagrangian}\label{Lag}
For strong interactions, the most general effective Lagrangian of 
$\chi$PT$_\sigma$ is of the form 
\begin{equation} 
\mathcal{L}^{}_{\mbox{\small $\chi$PT$_\sigma$}}
 =\ :\mathcal{L}^{d=4}_\mathrm{inv} 
 + \mathcal{L}^{d>4}_\mathrm{anom} + \mathcal{L}^{d<4}_\mathrm{mass}: \,,
\label{Lagr}\end{equation}
where 
\begin{equation}
  d_\mathrm{anom} = 4 + \gamma_{G^2}(\alpha_s) \qquad \mbox{and} 
  \qquad  d_\mathrm{mass} = 3 - \gamma_m(\alpha_s) 
\end{equation}
are the respective scaling dimensions of $G_{\mu\nu}^aG^{a\mu\nu}$ and 
$\bar{q}q$.  In lowest order (LO) of the chiral-scale expansion, we have 
$\gamma_m = \gamma_m(\alpha^{}_\mathrm{IR})$ and 
\begin{align}
  \gamma_{G^2}(\alpha_s) &\equiv \beta'(\alpha_s) - \beta(\alpha_s)/\alpha_s
  = \beta'(\alpha^{}_\mathrm{IR}) + O(\alpha_s-\alpha^{}_\mathrm{IR})\,,
\end{align}
so the resulting terms in (\ref{Lagr}) are 
\begin{align}
\mathcal{L}^{d=4}_\mathrm{inv,\,LO}
 &= \{c_{1}\mathcal{K} + c_{2}\mathcal{K}_\sigma 
     + c_{3}e^{2\sigma/F_\sigma} \} e^{2\sigma/F_\sigma} \,,
\notag \\[1mm]
\mathcal{L}^{d>4}_\mathrm{anom,\,LO}
&= \{(1-c_{1})\mathcal{K} + (1-c_{2})\mathcal{K}_\sigma 
      + c_4 e^{2\sigma/F_\sigma} \} e^{(2+\beta')\sigma/F_\sigma} \,,
\notag \\[1mm]
\mathcal{L}^{d<4}_\mathrm{mass,\,LO} 
 &= \mathrm{Tr}(MU^{\dagger}+UM^{\dagger})e^{(3-\gamma_m)\sigma/F_\sigma} \,,
\label{Lstr}
\end{align}
where 
\begin{equation}
  {\cal K} = \tfrac{1}{4}F_\pi^2\mathrm{Tr}(\partial_\mu U \partial^\mu U^\dagger) 
   \qquad \mbox{and} \qquad
  {\cal K}_\sigma = \tfrac{1}{2}(\partial_\mu\sigma)^2 \,.
\end{equation}
As $\alpha_s \to \alpha^{}_\mathrm{IR}$, the gluonic anomaly vanishes, so 
$\mathcal{L}_\mathrm{anom} = O(\del^2,M)$ and we must set $c_4 = O(M)$. Vacuum 
stability in the $\sigma$ direction about $\sigma = 0$ (no tadpoles) implies
\begin{align}
4c_3 + (4+\beta')c_4 &= - (3-\gamma_{m})\bigl\langle\mathrm{Tr}
        (MU^{\dagger}+UM^{\dagger})\bigr\rangle_{\mathrm{vac}} \notag \\
 &= - (3-\gamma_{m})F_\pi^2\bigl(m_K^2 + \tfrac{1}{2}m_\pi^2\bigr)\,,
\label{stable}\end{align}
so $c_3$ is also $O(M)$. Expanding (\ref{Lstr}) about $\sigma=0$ 
and $U=I$ yields the $\sigma\pi\pi$ coupling 
\begin{equation}
\mathcal{L}_{\sigma\pi\pi}
 = \bigl\{\bigl[2+(1-c_1)\beta'\bigr]|\del\bm{\pi}|^2 
    - (3 - \gamma_{m})m_\pi^2|\bm{\pi}|^2\bigr\}\sigma/(2F_\sigma) \,,
    \label{Lsigpi}
\end{equation}
while the corresponding $\sigma\pi\pi$ vertex for an on-shell dilaton is
\begin{align}
g_{\sigma\pi\pi} = 
 -\frac{1}{2F_\sigma}\Big\{\bigl[2+(1-c_1)\beta'\bigr]m_\sigma^2 
 + 2 \bigl[1-\gamma_m-(1-c_1)\beta'\bigr]m_\pi^2 \Big\}\,.
\label{on-shell}
\end{align}  

\section{Effective Energy-Momentum Tensor and its Trace}\label{tensor}
In any field theory, the energy-momentum tensor can be identified by 
adding a gravitational source field $g_{\mu\nu}(x)$ coupled to matter 
fields in a generally covariant fashion.  In $\chi$PT$_\sigma$, this 
amounts to the substitution
\begin{equation}
  {\cal L}_{\chi\mathrm{PT}_\sigma}[U,U^\dagger,\sigma] \to 
  {\cal L}_{\chi\mathrm{PT}_\sigma}[U,U^\dagger,\sigma,g_{\mu\nu}]\,,
\end{equation}
where the new effective Lagrangian must be constructed in terms of generally 
covariant operators. Then the energy-momentum tensor is defined via the 
variation
\begin{equation}
  \theta_{\mu\nu}(x) =  2\left[\frac{\delta}{\delta g^{\mu\nu}(x)} 
        \sqrt{-g} {\cal L}[U,U^\dagger,\sigma,g_{\mu\nu}]
        \right]_{g_{\mu\nu}=\eta_{\mu\nu}}\,,
\end{equation}
where $g=\mathrm{det}(g_{\mu\nu})$ is the determinant of the metric tensor and 
$\eta_{\mu\nu}$ is the flat Minkowski metric. Generalising Donoghue and 
Leutwyler~\cite{Don91}, we obtain the lowest order result
\begin{align}
  \theta_{\mu\nu} =\ &\big[ 
  \tfrac{1}{2}F_\pi^2\mathrm{Tr}\big(\partial_\mu U\partial_\nu U^\dagger\big) 
  -g_{\mu\nu}{\cal K} \big] 
  \big[ c_1 e^{2\sigma/F_\sigma} + (1-c_1) e^{(2+\beta')\sigma/F_\sigma} \big] 
  \nn \\
  &+\big(\partial_\mu\sigma\partial_\nu\sigma - g_{\mu\nu} {\cal K}_\sigma \big) 
  \big[ c_2e^{2\sigma/F_\sigma} + (1-c_2) e^{(2+\beta')\sigma/F_\sigma}\big] 
  \nn \\
  &-g_{\mu\nu}\mathrm{Tr}\big( MU^\dagger + UM^\dagger \big)e^{(3-\gamma_m)\sigma/F_\sigma} 
  -g_{\mu\nu}e^{4\sigma/F_\sigma}\big( c_3 + c_4e^{\beta'\sigma/F_\sigma} \big)\,.
  \label{can em tensor}
\end{align}
The trace of (\ref{can em tensor}) involves \textit{scale invariant} 
operators like 
$\mathrm{Tr}\big(\partial_\mu U\partial^\mu U^\dagger\big)e^{2\sigma/F_\sigma}$ 
which obscure the connection between the scale invariance and a conserved 
dilatation current $D_\mu$.  To remedy this, we ``improve'' 
$\theta_{\mu\nu}$~\cite{CCJ} by adding a term
\begin{equation}
  I_{\mu\nu} = 
  \frac{F_\sigma^2}{6}(g_{\mu\nu}\partial^2 - \partial_\mu\partial_\nu) 
  \left[ c_2 e^{2\sigma/F_{\sigma}} 
  + \frac{2(1-c_2)}{2+\beta'}e^{(2+\beta')\sigma/F_\sigma} \right]\,,
  \label{imp}
\end{equation}
such that the trace of
\begin{equation}
  \theta_{\mu\nu} \big|_\mathrm{eff} = \theta_{\mu\nu} + I_{\mu\nu}\,,
\end{equation}
is given entirely in terms of explicit scale-breaking operators ${\cal L}_d$ 
of scale dimension $d$:
\begin{equation}
  \partial^\mu D_\mu|_\mathrm{eff} = \theta_\mu^\mu \big|_\mathrm{eff} 
                                 = \sum_d (d-4){\cal L}_d \,.
  \label{trace req}
\end{equation}

Explicitly, the improved trace is
\begin{align} 
  \theta_\mu^\mu \big|_\mathrm{eff} =\ &\beta' {\cal L}_\mathrm{anom}^{d>4} 
  - (1+\gamma_m){\cal L}_\mathrm{mass}^{d<4} \nn \\
  =\ &\beta' \bigl\{(1-c_{1})\mathcal{K} + (1-c_{2})\mathcal{K}_\sigma
      + c_4 e^{2\sigma/F_{\sigma}}\bigr\}e^{(2+\beta')\sigma/F_{\sigma}}  \nn \\
 &-(1+\gamma_m) \mathrm{Tr}(MU^{\dagger}+UM^{\dagger})e^{(3-\gamma_{m})\sigma/F_{\sigma}}\,.
  \label{em trace}
\end{align}
It vanishes in the chiral-scale limit (\ref{scale}) only if the low-energy 
constants associated with $d>4$ operators satisfy
\begin{equation}
  c_1 = c_2 = 1 \,, \qquad \mbox{for } m_{u,d,s} \to 0 \mbox{ and } 
  \alpha_s \to \alpha^{}_\mathrm{IR}\,,  \label{c1eqc2}
\end{equation}
in addition to the condition $c_4=O(M)$ required by tadpole cancellation 
(\ref{stable}). Note that the condition $c_1 \to 1$ in (\ref{c1eqc2}) ensures that
chiral currents have vanishing anomalous dimensions. We can summarise these LO
conditions by writing  
\begin{equation}
  c_i = 1 + O(M) \,, \qquad i=1,2\,,
  \label{constraint}
\end{equation}
where the $O(M)$ term is a linear superposition of $O(p^2,M)$ operators and 
associated LECs.

\section{Improved Predictions}\label{predict}
An immediate consequence of the constraint (\ref{constraint}) is that the 
$\sigma \pi\pi$ coupling for an on-shell dilaton (\ref{on-shell}) takes a 
particularly simple form
\begin{equation}
g_{\sigma \pi\pi} = 
- \frac{1}{F_\sigma}\big[m_\sigma^2 + (1-\gamma_m)m_\pi^2 \big]\,, 
\qquad \mbox{where } -1 \leq 1-\gamma_m < 2\,.
\label{gsig simple}
\end{equation}
Since the narrow-width approximation is valid in lowest order 
$\chi$PT$_\sigma$~\cite{CT2}, we have
\begin{equation}
\Gamma_{\sigma\pi\pi} = 
\frac{|g_{\sigma \pi\pi}|^2}{16\pi m_\sigma}\sqrt{1-4m_\pi^2/m_\sigma^2}\,,
\end{equation}
and this allows us to obtain bounds on $F_\sigma$ from dispersive analyses of 
$\pi\pi$ scattering based on the Roy equations.  For example, the $f_0/\sigma$'s 
mass and width from~\cite{Cap06}
\begin{equation}
m_\sigma = 441^{+16}_{-8} \mbox{ MeV} \,, \qquad 
\Gamma_{\sigma\pi\pi} = 544^{+18}_{-25} \mbox{ MeV}\,,
\end{equation}
constrain $F_\sigma$ to lie within the interval 
$44 \mbox{ MeV} \leq F_\sigma \leq 61 \mbox{ MeV}$, where we have allowed 
$1-\gamma_m$ to vary according to (\ref{gsig simple}). For the moment, 
we assume that NLO corrections are not a problem.

With $F_\sigma$ fixed in this manner, we can now use the Golberger-Treiman 
relation for dilatons~\cite{MGM62} to \emph{predict} the value for the $\sigma NN$ coupling.  
We find $16 \leq g_{\sigma NN} \leq 21$, which is somewhat larger than previous 
phenomenological determinations~\cite{CC08,CC10}. Another important application 
concerns $\sigma\to\gamma\gamma$, where an analysis~\cite{CT1,CT2} of the 
electromagnetic trace anomaly in $\chi$PT$_\sigma$ relates the 
$\sigma\gamma\gamma$ coupling to (\ref{RIR}):
\begin{equation}
g_{\sigma\gamma\gamma} = 
\frac{2\alpha}{3\pi F_\sigma} \Big( R_\mathrm{IR} - \tfrac{1}{2} \Big)\,.
\end{equation}
By fixing $g_{\sigma\gamma\gamma}$ from the di-photon width 
$\Gamma_{\sigma\gamma\gamma} = 2.0 \pm 0.2\mbox{ keV}$~\cite{Hof11}, we find 
$2.4 \leq R_\mathrm{IR} \leq 3.1$, which is to be compared with our previous 
estimate $R_\mathrm{IR}\approx 5$~\cite{CT1,CT2}.

\section{Proposal to test $K \to \pi$ on the Lattice}\label{lattice}
The key idea \cite{RJC86} is to keep both $K$ and $\pi$ on shell and allow 
$O(m_K)$ momentum transfers.

The lowest-order diagrams for the decay $K \to \pi\pi$ in 
Fig.~\ref{fig:beta chpt} (B) are derived from an effective weak 
$\chi$PT$_\sigma$ Lagrangian \cite{CT1,CT2}
\begin{align}
{\cal L}_{\mathrm{weak}} = Q_{8}\sum_n g_{8n}e^{(2 -\gamma_{8n})\sigma/F_\sigma}
+ g_{27}Q_{27}e^{(2-\gamma_{27})\sigma/F_\sigma}
+ Q_{mw}e^{(3-\gamma_{mw})\sigma/F_\sigma} + \mbox{h.c.}
\label{weak}\end{align}
which reduces to the standard $\chi$PT$_3$ Lagrangian
\begin{equation}
\left.\mathcal{L}_{\mathrm{weak}}\right|_{\sigma=0} 
= g_{8}Q_{8} + g_{27}Q_{27} + Q_{mw} + \mathrm{h.c.}\
\label{usual}\end{equation}
in the limit $\sigma \to 0$. Eqs.~(\ref{weak}) and (\ref{usual}) contain 
an octet operator \cite{Cro67}
\begin{equation}
Q_{8} = {J}^\mu_{13}{J}^{}_{\mu 21} - {J}^\mu_{23}{J}^{}_{\mu 11}
\ , \quad
{J}^\mu_{ij} = (U\partial^{\mu}U^{\dagger})_{ij}
\label{eqn:octet}
\end{equation}
the $U$-spin triplet component \cite{RJC86,Gaill74} of a \textbf{27} operator
\begin{equation}
Q_{27} = {J}^\mu_{13}{J}^{}_{\mu 21} + \tfrac{3}{2}{J}^\mu_{23}{J}^{}_{\mu 11}
\label{eqn:27-plet}
\end{equation}
and a weak mass operator \cite{Bern85}
\begin{equation}
Q_{mw} = \mathrm{Tr} (\lambda_6 - i\lambda_7)
      \bigl(g_MMU^\dagger + \bar{g}_MUM^\dagger\bigr) \,.
\label{eqn:weak mass}
\end{equation}
Powers of $e^{\sigma/F_\sigma}$ are used to adjust the operator dimensions of
$Q_8,\, Q_{27},$ and $Q_{mw}$ in (\ref{weak}), with octet quark-gluon 
operators allowed to have differing dimensions at $\alpha^{}_\mathrm{IR}$.

In 1985, it was observed \cite{RJC86}  that the isospin-$\frac{1}{2}$ 
term $Q_{mw}$ in Eq.~(\ref{usual}), when combined with the strong mass term, 
would be removed by vacuum realignment and therefore could not help solve the 
$\Delta I = 1/2$ puzzle. In $\chi$PT$_\sigma$, the outcome is different 
\cite{CT1,CT2} due to the $\sigma$ dependence of the $Q_{mw}$ term in 
Eq.~(\ref{weak}). Provided there is a mismatch between the weak mass 
operator's dimension $(3-\gamma_{mw})$ and the dimension $(3-\gamma_{m})$ 
of ${\cal L}_\mathrm{mass}$, the $\sigma$ dependence of $Q_{mw} e^{(3-\gamma_{mw})/F_\sigma}$ 
cannot be eliminated by a chiral rotation. As a result, there is a residual 
interaction $\mathcal{L}_{K_S\sigma} = g^{}_{K_S\sigma}K_{S}\sigma$
which mixes $K_{S}$ and $\sigma$ in \emph{lowest} $O(p^2)$ order%
	\footnote{We have corrected a factor of 2 in the formula
	for the $K_S\sigma$ coupling in our original papers~\cite{CT1,CT2}.}
\begin{equation}
g^{}_{K_S\sigma} 
 = (\gamma_{m} - \gamma_{mw})\mathrm{Re}\{(2m^2_K - m^2_\pi)\bar{g}_M
                    - m^2_\pi g_M\}F_{\pi}/F_{\sigma} 
\end{equation}
and produces the $\Delta I = 1/2$ $\sigma$-pole amplitude of Fig.~\ref{fig:beta chpt} (B).

The $\chi$PT$_3$ analysis of 1985 \cite{RJC86} included a suggestion that
kaon decays be tested by applying lattice QCD to the weak process $K \to 
\pi$, with \emph{both} $K$ and $\pi$ on shell. It was made at a time when 
low-lying scalar resonances ($\epsilon(700)$ before 1974, $f_0(500)$ since 
1996) were thought not to exist. 

This proposal now needs to be taken seriously because:
\begin{itemize}
\item Lattice calculations are much easier with only two particles on shell 
instead of the three in $K \to\pi\pi$ (all on shell) being analysed by the 
RBC/UKQCD collaborations \cite{RBC1,RBC2}.
\item The 1985 analysis is easily extended to $\chi$PT$_\sigma$ by 
including $\sigma/f_0$ pole amplitudes in chiral Ward identities connecting 
on-shell $K \to \pi\pi$ to $K \to \pi$ on shell. The no-tadpoles theorem 
\begin{equation}
\langle K | \mathcal{H}_{\mathrm{weak}} |\mathrm{vac}\rangle
= O\bigl(m_s^2 - m_d^2\bigr)\,, \ K\mbox{ on shell}\,,
\label{tadpole}
\end{equation}
remains valid.
\item The lattice result for $K \to\pi\pi$ on-shell will not distinguish
$\Delta I = 1/2$ contributions from the $g_8$ contact diagram and the
$\sigma/f_0$ pole diagram in Fig.~\ref{fig:beta chpt} (B). A lattice
calculation of  $K \to \pi$ on shell would measure $g_8$ (and $g_{27}$) 
directly, with no interference from $\sigma/f_0$ poles. Then we would
\emph{finally} learn whether $g_8$ is unnaturally large or not.
\end{itemize}
A key feature of the proposal is that the operator in the on-shell
amplitude $\langle\pi|[F_5,\mathcal{H}_{\mathrm{weak}}]|K\rangle$ 
necessarily carries \emph{non-zero} momentum $q^\mu = O(m_K)$. 
For either $\chi$PT$_\sigma$ or $\chi$PT$_3$, the $K \to \pi$ amplitude 
can be evaluated in the range 
\begin{equation}
- m_K^2 \lesssim q^2 \leqslant \bigl(m_K - m_\pi\bigr)^2  \,.
\end{equation}

We highlight the point $q^\mu \not= 0$  because since 1985, there has been 
a widespread misconception in the literature%
\footnote{We thank the final referee of our long paper \cite{CT2} for 
drawing our attention to this.}
that the analysis \cite{RJC86} involved setting $q^\mu = 0$ as in 
\cite{Bern85}, with the pion in $K \to \pi$ sent off shell via an
interpolating operator. There was and is no reason for this. For example, 
when writing a soft meson theorem for $\Sigma \to p\pi$, it is not 
necessary to force one of the baryons off shell.

A lattice determination of the on-shell  $K\to\pi$ amplitude is also essential 
for dispersive analyses~\cite{Buch01,Col01} where ${\cal H}_\mathrm{weak}$ 
is allowed to carry non-zero momentum in $K\to\pi\pi$. The key observation of
this work is that this momentum can be chosen such that only one soft pion is 
needed: the extrapolated $K\to\pi\pi$  amplitude and on-shell $K\to\pi$ amplitude
are related up to chiral corrections which are only $O(m_\pi^2)$.
 
\section{Issues}\label{issues}

When considering the validity of $\chi$PT$_\sigma$, it is important to
avoid any presumption that dimensional transmutation necessarily implies
that $\theta^\mu_\mu$ is large and $\not= 0$. Implicit in this 
intuition is a prejudice that scale invariance cannot be strongly
broken via the vacuum when $\theta^\mu_\mu \to 0$. If the dilaton is a 
true NG boson, i.e.\  $m_\sigma \to 0$ with $F_\sigma \not= 0$ for 
$\theta^\mu_\mu \to 0$, it can couple to mass insertion terms in 
Callan-Symanzik equations and cause them to be \emph{non-zero} in the 
zero-mass limit. Then Green's functions do not exhibit the power-law scaling 
expected for manifestly scale-invariant field theories.

This point is illustrated for the quark condensate in 
Fig.~\ref{fig:beta chpt} (A). In scenario 1 (top diagram), the
running of $\alpha_s$ is driven by the presence of quantities like
$\langle \bar q q \rangle_\mathrm{vac}$ (a mechanism often cited in
papers on walking gauge theories~\cite{Del10}). In scenario 2 (bottom 
diagram), the running coupling freezes at $\alpha^{}_\mathrm{IR}$, where 
the condensate is a \emph{scale-breaking property of the vacuum}.

Lattice investigations of IR fixed points inside the conformal window 
$8 \lesssim N_f \leqslant 16$ all depend on naive scaling of Green's 
functions \cite{Del10}, so they correspond to \emph{scale-invariant 
vacua}. A recent lattice study \cite{lattice} of the running of 
$\alpha_s$ for two flavors with \emph{no} naive scaling suggests that it 
freezes: the fixed point realises scale invariance in NG mode, i.e.\
with a scale-breaking vacuum. That is what $\chi$PT$_\sigma$ assumes
for three flavors.

The term ``dilaton'' often refers to a spin-0 particle or resonance 
which couples to $\theta_{\mu\nu}$ and acquires its mass ``spontaneously''
due to self interactions. Originally, this idea concerned a scalar component of 
gravity \cite{Fujii}, but now it is a key ingredient of dynamical electroweak 
symmetry breaking (pp.~198 and 1622-3, PDG tables~\cite{PDG}). This 
approximates theories with \emph{scale-invariant vacua}, as is evident in 
walking technicolor. Therefore it has \emph{nothing} to do with our 
dilaton~\cite{Carr}.

It is well known that a resonance cannot be represented by a local
interpolating operator, so is the fact that $\sigma/f_0(500)$ has a 
finite width a problem for $\chi$PT$_\sigma$? The answer is ``no''
because $\chi$PT$_\sigma$ is an expansion in powers and logarithms
of $m_{\pi,K,\eta,\sigma}$ with coefficients determined in the \emph{exact}
chiral-scale limit (\ref{scale}) where $\sigma$ has zero width \cite{CT2}. 
In any perturbation theory, decay rates are calculated that way.

A related remark concerns what is current best practice for scenario 1.
The resonance $f_0(500)$ is treated as a member of the non-NG sector 
with an accidentally small mass. It causes $\chi$PT$_3$ to produce 
divergent expansions for amplitudes involving $f_0(500)$ poles:
the radius of convergence is too small. Instead, these amplitudes
are approximated dispersively via contributions from the dominant 
$f_0(500)$ poles with corrections from nearby thresholds, subject to 
exact chiral $SU(3)\times SU(3)$ constraints such as Adler zeros. One 
would certainly not use local fields in this framework.

However $\chi$PT$_\sigma$ is a more ambitious theory. Having promoted
$\sigma/f_0$ to the NG sector, we expect convergent asymptotic expansions 
for \emph{all} mesonic amplitudes (scenario 2). The NLO corrections are still 
being worked out, but a first guess is to set all multi-dilaton vertices to zero. 
That is equivalent to adding the simplest dilaton diagrams to all $\chi$PT$_3$ 
diagrams. It seems to produce amplitudes very similar to those of the dispersive 
approximations of scenario 1. 

\section{Acknowledgements}
We thank Claude Bernard, Gilberto Colangelo, Gerhard Ecker, Maarten Golterman, 
Martin Hoferichter, Heiri Leutwyler, and Daniel Phillips for useful comments 
about $\chi$PT$_\sigma$ and the work we presented at CD2015.  We also 
thank Nicolas Garron for informative discussions regarding RBC/UKQCD's 
analyses of $K\to \pi\pi$. LCT is supported by the Swiss National Science 
Foundation.



\begin{thebibliography}{99}
\bibitem{Adler77} S.~L.~Adler, J.~C.~Collins and A.~Duncan, 
\emph{Energy-Momentum-Tensor Trace Anomaly in Spin 1/2 Quantum Electrodynamics},
Phys.\ Rev.\ D \textbf{15}, 1712 (1977).

\bibitem{Mink76}  P.~Minkowski,
\emph{On the Anomalous Divergence of the Dilatation Current in Gauge Theories}, 
Berne PRINT-76-0813, September 1976.

\bibitem{Niels77} N.~K.~Nielsen,
\emph{The Energy Momentum Tensor in a Nonabelian Quark Gluon Theory},
Nucl.\ Phys.\ \textbf{B120}, 212 (1977).

\bibitem{Coll77} J.~C.~Collins, A.~Duncan and S.~D.~Joglekar, 
\emph{Trace and Dilatation Anomalies in Gauge Theories}, 
Phys.\ Rev.\  D \textbf{16}, 438 (1977).

\bibitem{CT1} 
R.~J.~Crewther and L.~C.~Tunstall,
\emph{Origin of $\Delta I=1/2$ Rule for Kaon Decays: QCD Infrared Fixed Point},
{\tt arXiv:1203.1321}.

\bibitem{CT2}
R.~J.~Crewther and L.~C.~Tunstall,
\emph{$\Delta I=1/2$ rule for kaon decays derived from QCD infrared fixed point},
Phys.\ Rev.\ D {\bf 91}, 034016 (2015)
[{\tt arXiv:1312.3319}].

\bibitem{Cap06} I.~Caprini, G.~Colangelo and H.~Leutwyler, 
  \emph{Mass and width of the lowest resonance in QCD}, 
  Phys.\ Rev.\ Lett.\ \textbf{96}, 132001 (2006)
  [{\tt arXiv:hep-ph/0512364}].

\bibitem{Kam11}  R.~Garc\'{i}a-Mart\'{i}n, R.~Kami\'{n}ski, J.~R.~Pel\'{a}ez
  and J.~R. de~Elvira, 
  \emph{Precise determination of the $f_0(600)$ and $f_0(980)$ pole parameters 
  from a dispersive data analysis}, 
  Phys.\ Rev.\ Lett.\  {\bf 107}, 072001 (2011)
  [{\tt arXiv:1107.1635}].

\bibitem{PDG}
  K.~A.~Olive {\it et al.} [Particle Data Group Collaboration],
  \emph{Review of Particle Physics}, 
  Chin.\ Phys.\ C {\bf 38}, 090001 (2014).

\bibitem{MGM62}
  M.~Gell-Mann, 
  \emph{Symmetries of Baryons and Mesons},
  Phys.\ Rev.\ {\bf 125}, 1067 (1962), footnote 38.

\bibitem{CC08} 
  A.~Calle Cordon and E.~Ruiz Arriola,
  \emph{Scalar meson mass from renormalized One Boson Exchange Potential}, 
  AIP Conf.\ Proc.\  {\bf 1030}, 334 (2008)
  [{\tt  arXiv:0804.2350}].
  
\bibitem{CC10} 
  A.~Calle Cordon and E.~Ruiz Arriola,
  \emph{Renormalization vs Strong Form Factors for One Boson Exchange Potentials}, 
  Phys.\ Rev.\ C {\bf 81}, 044002 (2010)
  [{\tt arXiv:0905.4933}].
  
\bibitem{RJC86} 
  R.~J.~Crewther,
  \emph{Chiral Reduction of $K \to 2 \pi$ Amplitudes},
  Nucl.\ Phys.\ B {\bf 264}, 277 (1986).

\bibitem{Don91} 
  J.~F.~Donoghue and H.~Leutwyler,
  \emph{Energy and momentum in chiral theories},
  Z.\ Phys.\ C {\bf 52}, 343 (1991).

\bibitem{CCJ} 
  C.~G.~Callan, Jr., S.~R.~Coleman and R.~Jackiw,
  \emph{A new improved energy-momentum tensor}, 
  Annals Phys.\  {\bf 59}, 42 (1970).
  
\bibitem{Hof11} 
  M.~Hoferichter, D.~R.~Phillips and C.~Schat,
  \emph{Roy-Steiner equations for $\gamma\gamma\to\pi\pi$},
  Eur.\ Phys.\ J.\ C {\bf 71}, 1743 (2011)
  [{\tt arXiv:1106.4147}].

\bibitem{Cro67} 
  J.~A.~Cronin, 
  \emph{Phenomenological Model of Strong and Weak Interactions in Chiral 
  $U(3)\times U(3)$},
  Phys.\ Rev.\ \textbf{161}, 1483 (1967).

\bibitem{Gaill74} M.~K.~Gaillard and B.~W.~Lee, 
  \emph{$\Delta I = 1/2$ Rule for Nonleptonic Decays in Asymptotically 
  Free Field Theories},
  Phys.\ Rev.\ Lett.\ \textbf{33}, 108 (1974).

\bibitem{Bern85} C.~Bernard, T.~Draper, A.~Soni, H.~D.~Politzer 
  and M.~B.~Wise, 
  \emph{Application of chiral perturbation theory to $K \to 2\pi$ decays},
  Phys.\ Rev.\ D \textbf{32}, 2343 (1985).

\bibitem{RBC1} 
  P.~A.~Boyle \emph{et al.} [RBC/UKQCD Collaboration],
  \emph{Emerging understanding of the $\Delta I = 1/2$ Rule from Lattice QCD},
  Phys.\ Rev.\ Lett.\ {\bf 110}, 152001 (2013)	
  [{\tt arXiv:1212.1474}].

\bibitem{RBC2}
  Z. Bai \emph{et al.} [RBC/UKQCD Collaboration],
  \emph{Standard-model prediction for direct $CP$ violation in 
  $K \to \pi\pi$ decay},
  {\tt arXiv:1505.07863}.

\bibitem{Buch01} 
  M.~Buchler, G.~Colangelo, J.~Kambor and F.~Orellana,
  \emph{Dispersion relations and soft pion theorems for $K\to \pi\pi$},
  Phys.\ Lett.\ B {\bf 521}, 22 (2001)
  {\tt [arXiv:hep-ph/0102287]}.

\bibitem{Col01} 
  G.~Colangelo,
  \emph{Chiral perturbation theory, dispersion relations and final state 
  interactions in $K\to\pi\pi$},
  Nucl.\ Phys.\ Proc.\ Suppl.\  {\bf 106}, 53 (2002)
  {\tt [arXiv:hep-lat/0111003]}.

\bibitem{Del10}
  L.~Del~Debbio, 
  \emph{The conformal window on the lattice},
  Proc.\ Sci.\  LATTICE2010 (2010) 004
  [{\tt	arXiv:1102.4066}].
 
\bibitem{lattice}
  R.~Horsley, H.~Perlt, P.~E.~L.~Rakow, G.~Schierholz and A.~Schiller,
  \emph{The $SU(3)$ beta function from numerical stochastic perturbation
  theory},
  Phys.\ Lett.\ B {\bf 728}, 1 (2014)
  [{\tt arXiv:1309.4311}].

\bibitem{Fujii} Y.~Fujii, 
  \emph{Dilaton and Possible Non-Newtonian Gravity},
  Nat.\ Phys.\ Sci.\ {\bf 234}, 5 (1971).

\bibitem{Carr} P.~Carruthers, 
  \emph{Broken scale invariance in particle physics},
  Phys.\ Rep.\ C {\bf 1}, 1 (1971). 

\end{thebibliography}
\end{document}